# Nanometer displacement measurement based on metrological self-mixing grating interferometer traceable to the pitch standard of one-dimension chromium self-traceable grating

ZHENJIE GU,[1,2,3,4,5] ZHANGNING XIE,[1,2,3,4,5] ZHIKUN CHANG,[1,2,3,4,5] GUANGXU XIAO[1,2,3,4,5], ZHIJUN YIN[1,2,3,4,5], ZICHAO LIN[1,2,3,4,5], ZHOU TONG[1,2,3,4,5], LIHUA LEI[6], TAO JIN[7], DONGBAI XUE[1,2,3,4,5], XIAO DENG[1,2,3,4,5],[*], XINBIN CHEN[1,2,3,4,5,*], AND TONGBAO LI[1,2,3,4,5]

[1] *Institute of Precision Optical Engineering, Tongji University, Shanghai 200092, China*
[2] *MOE Key Laboratory of Advanced Micro-Structured Materials, Tongji University, Shanghai 200092, China*
[3] *Shanghai Frontiers Science Center of Digital Optics, Tongji University, Shanghai 200092, China*
[4] *Shanghai Professional Technical Service Platform for Full-Spectrum and High-Performance Optical Thin Film Devices and Applications, Tongji University, Shanghai 200092,China*
[5] *School of Physics Science and Engineering, Tongji University, Shanghai 200092, China*
[6] *Shanghai Institute of Measurement and Testing Technology, Shanghai 200093, China*
[7] *University of Shanghai for Science and Technology, Shanghai 200093, China*

Correspondence:*18135@tongji.edu.cn;*

**Abstract:** Traceability of precision instrument and measuring method is the core issue in metrology science. In the field of nanometer length measurement, the laser interferometers are usually used to trace the measurement value to the laser wavelength, but the laser wavelength is sensitive to the environment disturbance. Chromium self-traceable grating is an ideal nanometer length reference grating with pitch traceability, fabricated by the atomic lithography technique. The new nanometer length traceability chain can be established based on the pitch traceability of chromium self-traceable grating, which is often used to calibrate the systematic error of the atomic force microscope. In this paper, the metrological self-mixing grating interferometer based on the chromium self-traceable grating (SMGI-Cr) is firstly established, whose interfere phase is traceable to the pitch of the chromium self-traceable grating directly and traceable to the chromium atomic transition frequency of energy level $^7S_3 \rightarrow ^7P_4$ indirectly. The nanometer displacement measurement is also achieved by the SMGI-Cr. The measurement error is no more than 0.2366%, compared to a commercial interferometer.

## 1. Introduction

Precision displacement measurement is a critical technology widely used in quality control in precision manufacturing fields such as semiconductor industry and nanomaterial preparation. The traceability of measurement is one of the core issues in the field of metrology science, which ensures the accuracy and consistency of measurement results. The accuracy of precision displacement measurement structures also requires traceability to ensure. According to the definition of ISO9000 of The so-called traceability refers to the characteristic of "connecting measurement results or values of measurement standards with specified reference standards, usually national or international measurement standards, through an uninterrupted chain of comparison with specified uncertainty".

With the development of science and technology, the seven SI basic units have transformed from physical benchmarks to natural attributes with quantum properties. In 1999, the International Bureau of Metrology published a notice on the implementation of the definition of "meter", providing twelve international standard spectral lines for achieving the definition of "meter". It has become an international consensus to use laser interferometers to trace the length of "meters" to atomic or molecular spectral lines.

Laser self-mixing interferometer[1] is a dual beam interferometric measurement method based on the principle of optical feedback. It involves the feedback light from an external optical feedback element returning to the laser resonant cavity to interfere with the initial laser and establish a new steady-state self-mixing interference light field. The feedback light carries externally introduced phase information relative to the initial laser. When the phase changes, it will cause changes in the frequency and power of the self-mixing light field. The relationship between the phase change and the measured physical quantity can be constructed, and the measured physical quantity can be reconstructed through phase demodulation method. After more than half a century of development, laser self-mixing interferometers have been widely used in fields such as precision displacement measurement[2], absolute distance measurement[3], velocity measurement[4], and material refractive index measurement[5]. In the various applications, displacement measurement is the most basic application method of laser self-mixing interferometer, and other applications can be obtained based on the laser self-mixing displacement measurement system. According to the different feedback elements used in laser self-mixing interferometers, they can be divided into two types according to the traceability method. The first type uses a reflector as the optical feedback element, and its displacement and phase are associated by the laser wavelength, $\boldsymbol{\varphi = 4\pi\Delta x/\lambda}$. The measurement results are traceable to the laser wavelength, so a He-Ne laser light source is usually used to construct a laser self-mixing interferometer with metrological characteristics[6]. Another type is a self-mixing grating interferometer, which uses the $\boldsymbol{m}$-order diffraction light diffracted by a grating as the feedback light[7]. The displacement and phase are connected through the grating pitch $\boldsymbol{d}$, satisfying the $\boldsymbol{\varphi = 2\pi m\Delta x/d}$. Therefore, the measurement results can be traced back to the grating period value. Laser wavelength is sensitive to environmental disturbances and is easily affected by environmental temperature and air refractive index. For high-precision measurement, corresponding environmental compensation modules must be equipped. Compared to other methods, the grating pitch has better robustness to environmental changes, but its disadvantage is that the measurement range is limited by the grating area. At present, there are grating self-mixing displacement measurement schemes that combine helium neon lasers with transmission gratings[7], optical subdivision methods that use multiple diffraction to improve measurement resolution, and schemes that use multi-channel grating self-mixing interferometers to achieve two-dimensional[8] and three-dimensional displacement sensing[9]. It is worth noting that so far, all grating self-mixing interferometers have used general commercial holographic gratings, and their pitch values do not have traceability. For high requirements and standards such as semiconductor industry and nanomanufacturing, additional calibration is required for the grating pitch values or system measurement values.

Atomic lithography grating is a nanometer grating with pitch self-traceability characteristics prepared by laser focused atomic deposition technology[10]. By passing the collimated atomic beam through the vertical laser standing wave field, under the effect of the gradient force of the light field, the atoms are distributed on the substrate according to the intensity of the laser standing wave field to form a periodic grating structure. For one-dimensional atomic photolithography gratings, their pitch values can be traced back to half of the wavelength of the atomic transition energy level corresponding to the laser standing wave field. Jabez J. et al. from NIST in the United States first implemented a chromium atom lithography grating that can be used at room temperature using chromium atoms. They also demonstrated using laser diffraction method that the average pitch value of the one-dimensional chromium atom lithography grating prepared in the experiment is (212.7705 ± 0.0049) nm[11], and its pitch value can be directly traced back to the chromium transition energy $^7S_3 \rightarrow {}^7P_4$. Subsequently the research group from Tongji University in China also rendered the chromium atomic deposited grating as nanoscale ruler[12], and the structure is described carefully by using scanning electron microscope[13]. Therefore, chromium atom lithography grating is an ideal nanometer length reference material with self-traceability pitch, which can be used to directly calibrate error of nanometer measuring equipment such as atomic force microscope. Based on the self-traceability characteristics of chromium atom photolithography gratings, Deng Xiao et al. from Tongji University established a new type of nanometer length traceability chain and implemented a one-dimensional chromium self-traceability grating interferometer for nanometer displacement measurement[14].

This paper will combine the pitch self-traceability characteristics of chromium self-traceability gratings with the simple structure and high measurement accuracy of self-mixing interferometers to study a self-mixing grating interferometer based on chromium self-traceability grating feedback (SMGI-Cr), in order to achieve a metrological self-mixing precision displacement measurement method that can be directly applied to on-site measurement without additional calibration.

## 2. The experiment setup of SMGI-Cr

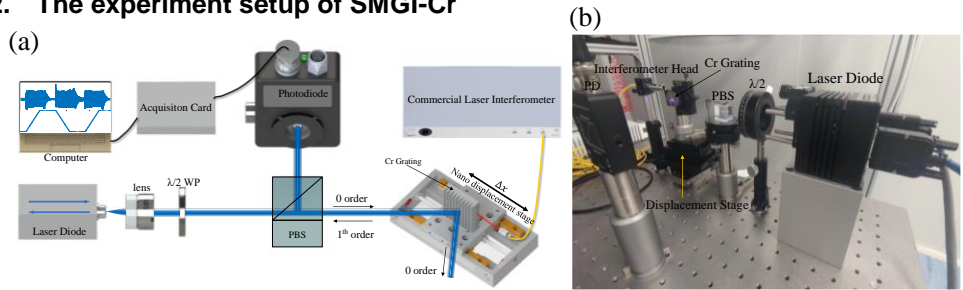

Fig1. (a) Experimental sketch map of SMGI-Cr (b)Experimental setup for SMGI-Cr

The system schematic of the self-mixing grating interferometer based on the Chromium self-traceable grating (SMGI-Cr) is described as follows: the optical source is radiated by a 405nm single-longitudinal-mode laser diode (Thorlabs，DL5146-101S), which is equipped with temperature control module and current control module. The output power is about 40mw. The focal length of the output laser is adjusted by an aspherical lens mounted on an adjusting frame and make the laser cover the area of the chromium self-traceability grating. The laser is splitted by a half-wave plate and a polarization beam splitter, which can divide the laser into reflected and transmitted light according to arbitrary ratio. The faint reflected laser is detected by a photodetector (Thorlabs，PDA25K2) to obtain the self-mixing power signal. The photodetector signal is collected by a data acquisition card(MCC 1608G-2AO) and processed by an computer. The transmitted laser is incident on the one-dimension chromium self-traceable grating whose pitch is $(212.78 \pm 0.1)$nm accurately. The profile of the chromium self-traceable grating is gaussian type and the height used in this experiment is 38nm with the area of 1mm×2mm. The incident angle is Littrow angle at which the one-order diffraction laser can coincide with the input laser and return to the laser diode inner cavity, generating self-mixing effect. The Littrow angle can be calculated by $\arcsin(\lambda/2d)$, which equals to 72.12 degree in this experiment. The chromium self-mixing grating is mounted on a one-dimension nanometer displacement stage driven by linear motor. Because of the weak diffraction efficiency no more than 5%, the SMGI-Cr normally operates under the weak feedback regime of self-mixing effect.

## 3. Displacement measurement principle of SMGI

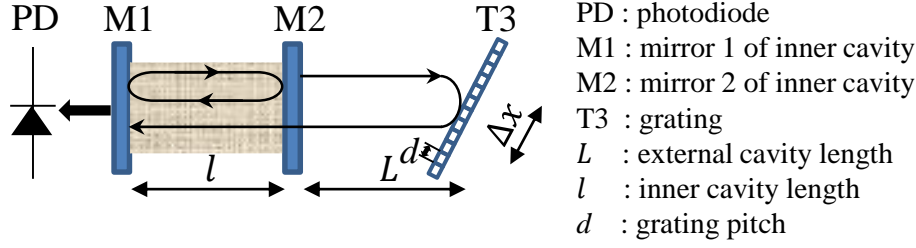

Fig2. The schematic diagram of three cavity model for the self-mixing grating interferometer

The theory model of SMGI can be explained by the three mirror cavity mode, which can be summarized into two equations, frequency equation and power equation:

$$\omega_0\tau = \omega\tau + C\sin(\omega\tau + \varphi(t) - \arctan\alpha)$$
$$P(t) = P_0 + P_{ac} = P_0[1 + m\cos(\omega\tau + \varphi(t))] = P_0[1 + m\cos(\varphi_F(t))]$$

Where $\omega$ and $\omega_0$ are laser angular frequency with and without feedback laser respectively, $\tau$ is the round-trip time in the external cavity, $C$ is the feedback enhancement factor, $\alpha$ is the linewidth enhancement factor, $P(t)$ and $P_0$ are the output power with and without feedback, m is the visibility of the signal, $\phi(t)$ is the phase modulation caused by the doppler shifting of grating. For a displacement $\Delta x$ moving along the grating's period direction, the m-order diffraction feedback laser will introduce a phase shift caused by the doppler frequency shift,

$$\varphi(t) = \frac{2\pi m \Delta x(t)}{d}$$

Where $\Delta x$ represent the displacement along the direction of the grating period, which has a linear relationship with the phase shift $\varphi(t)$. For real-time processing the data, the phase $\omega\tau$ caused by the frequency modulation of self-mixing effect is generally omitted and lead into a systematic periodic error item.

In the experiment, only the AC part of self-mixing signal $mP_0\cos(\varphi_F(t))$ is collected. In order to obtained the phase value $\varphi_F(t)$, the Hilbert transformation is adopted instead of solving the inverse cosine function of the self-mixing signal. For a real value signal $S(t)$, its Hilbert transformation is defined as

$$S_H(t) = H[S(t)] = \frac{1}{\pi}\int_{-\infty}^{+\infty}\frac{S(\tau)}{t-\tau}d\tau$$

The signal $S_H(t)$ will generate 90° phase shift relative to the original signal $S(t)$。And a $\cos(\varphi_F)$ input signal will transfer to $\sin(\varphi_F)$ signal after one-time HT. So the tangent function of $\varphi_F$ can be obtained, which is

$$\tan(\varphi_F) = \frac{S_H(t)}{S(t)} = \frac{H[\cos(\varphi_F)]}{\cos(\varphi_F)}$$

Combining solving the arctangent function and phase unwrapping algorithm[], $\varphi_F$ can be obtained.

$$\varphi_F = \arctan\left(\frac{S_H(t)}{S(t)}\right) + (-1)^n * k * \pi$$

Where $n$ is the number of the direction transition points, $k$ is the number of stripe. Omitting the $\omega\tau$ item means substituting $\varphi_F$ to $\varphi(t)$, then the displacement can be calculated by linear displacement-phase relationship.

## 4. Experiment Result and analysis

The collected original signal of SMGI-Cr is shown in figure 3(a), which is the ac signal part of the output optical intensity. The more distinct self-mixing signal extracted partially is presented in figure3(b). The original signal is composed of two parts, significant signal oscillating part and noisy signal part, which is caused by the actual moving mode of displacement stage is a trapezoidal mode. When the stage is paused, the self-mixing effect caused by the doppler frequency is disappeared and only the electronic noise is detected.

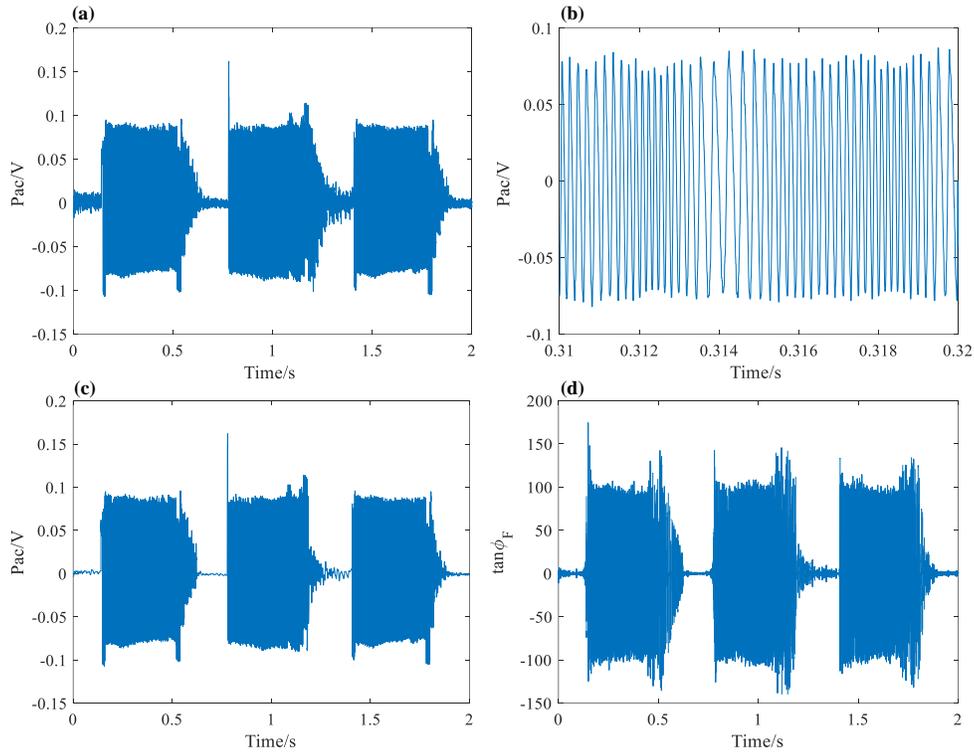

Fig3. (a)the original signal of self-mixing grating interferometer with chromium self-traceable grating feedback(SMGI-Cr), which is the ac part of the optical power represented by $P_{ac}$.(b) the partial extracted self-mixing signal from the original signal(c) the self-mixing signal with the noise filtered (d) the tangent value of phase $\varphi_F$

For eliminating the noise influence on the constructed displacement in the stationary area, the multiple moving average filtering is used to reducing the high-frequency noise, the filtered self-mixing signal is shown as figure 4(c),Then the sine signal is calculated from the imagine part of one-time HT signal. So the tangent value of phase is obtained by sine signal divided by original cosine signal, which is shown as figure 4(b).

Then the wrapped phase $\varphi_F$ is shown in figure 4(a), which is obtained directly by solving the inverse tangent function of the data in figure 3(d). Accumulating the wrapped phase directly without the direction description, the unwrapped phase is shown as figure 4(b).

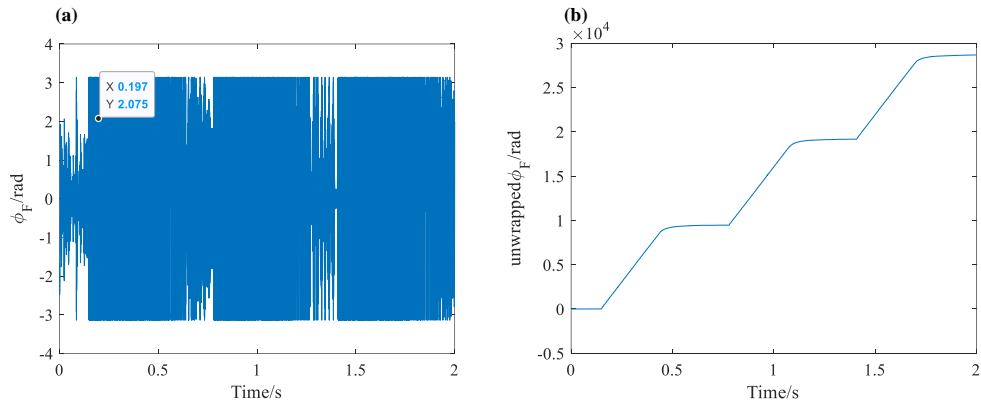

Fig4. (a)Wrapped phase picture of self-mixing signal (b) Unwrapped phase without setting the turning points

For getting the phase unwrapped phase, the moving direction must be determined. For heterodyne detection scheme, the changing of frequency difference relative to the static frequency is used to sensing[15]. For homodyne scheme, the tangent value of phase from the quadrature detection signal can be used to discern the direction[16]. For the moderate feedback, the direction is identified by the differentiation of the self-mixing because of the slope of signal is sharp enough and change sign after the turning point[17]. In this paper, the turning points are discriminated by searching the peaks from the reference displacement signal of a commercial Fabry-Perot interferometer(QuDIS). The displacement measured by the commercial interferometer is shown as figure 7(a), and all the turning points is marked by the red asterisk on the picture.

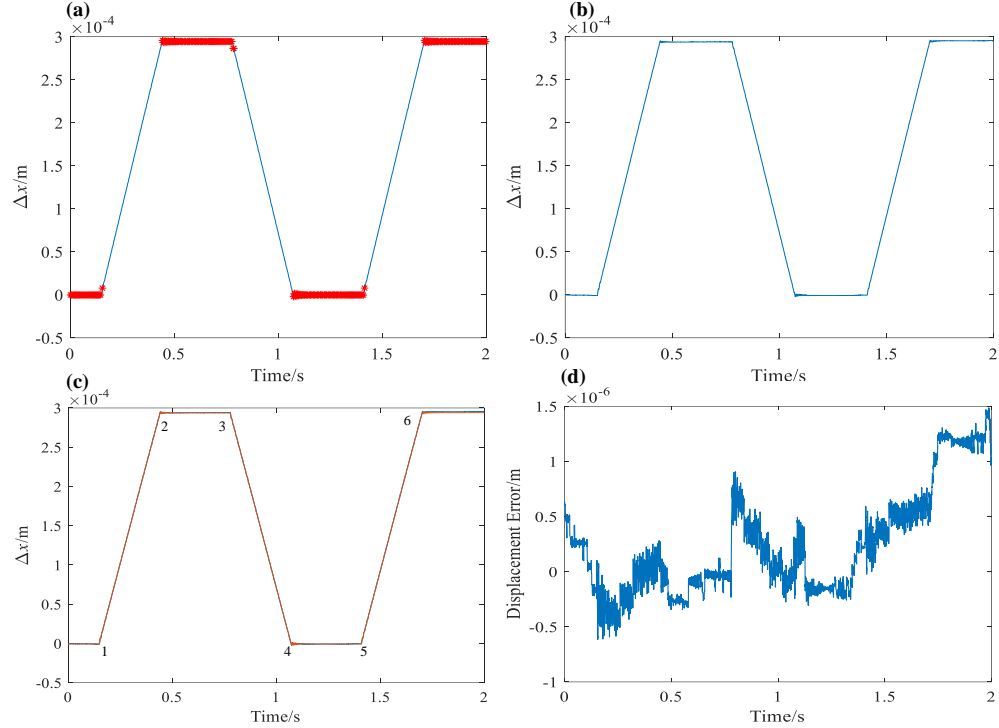

Fig7. (a)Displacement measured by the commercial Fabry-Perot interferometer, the turning points are marked by the red asterisk '*'.(b) Displacement measured by the SMGI-Cr after considering the turning points.(c) Displacement measured by the SMGI overlapping with the result from commercial interferometer (d) Displacement error calculated by displacement of SMGI-Cr minus that of commercial interferometer

Every time the moving direction reversed, the tangent value of self-mixing signal should change the sign. Considering the turning points and transferring unwrapped phase to displacement according to formula $\Delta x = \varphi d/2\pi$, where the self-traceable grating period $d$ is taken as 212.78nm, the displacement reconstructed from the SMGI-Cr is presented in figure 7(b).

In order to estimate the displacement measurement error between the SMGI-Cr and the commercial interferometer, the results of the displacement from the two interferometer are drew in one picture, shown as figure 7(c). The two displacement curves are almost overlapped, and their displacement error is exhibited in figure 7(d). The maximum displacement error is 1.4859*10$^{-6}$ nm. The local extremum near the turning area is chosen as the starting point and ending point of unidirectional displacement, shown in Table 1, whose corresponding time is (0.1495, 0.444, 0.7795, 1.074, 1.409, 1.704) respectively. The absolute value of relative error $|\delta|$, which equals $\left|(\delta x_{SMGI-Cr} - \delta x_{quDIS})/\delta x_{quDIS}\right|$, was no more than 0.2366%, listed in the table 2.

**Table 1 the position value at every turning point of SMGI-Cr（$\Delta x_{SMGI-Cr}$）, that of commercial laser interferometer QuDIS（$\Delta x_{quDIS}$）, and the displacement amplitude of one-way trip, $\delta x_{SMGI-Cr}$ and $\delta x_{quDIS}$**

| Index | $\Delta x_{quDIS}/m$ | Index | $\Delta x_{quDIS}/m$ | $\delta x_{quDIS}/m$ |
|---|---|---|---|---|
| 1 | -8.391×10⁻⁷ | 2 | 2.952×10⁻⁴ | 2.960391×10⁻⁴ |
| 3 | 2.942×10⁻⁴ | 4 | -2.367×10⁻⁶ | -2.96567×10⁻⁴ |
| 5 | -8.006×10⁻⁷ | 6 | 2.953×10⁻⁴ | 2.961006×10⁻⁴ |
| Index | $\Delta x_{SMGI-Cr}/m$ | Index | $\Delta x_{SMGI-Cr}/m$ | $\delta x_{SMGI-Cr}/m$ |
| 1 | -1.1×10⁻⁶ | 2 | 2.952×10⁻⁴ | 2.963×10⁻⁴ |
| 3 | 2.941×10⁻⁴ | 4 | -2.3×10⁻⁶ | -2.964×10⁻⁴ |
| 5 | 4×10⁻⁷ | 6 | 2.958×10⁻⁴ | 2.954×10⁻⁴ |

**Table 2 the displacement relative error between SMGI-Cr and commercial laser interferometer QuDIS**

| Index | $\delta x_{SMGI-Cr}/m$ | $\delta x_{quDIS}/m$ | $|\delta|$ |
|---|---|---|---|
| 1 | 2.963×10⁻⁴ | 2.960391×10⁻⁴ | 0.0881% |
| 2 | -2.964×10⁻⁴ | -2.96567×10⁻⁴ | 0.0563% |
| 3 | 2.954×10⁻⁴ | 2.961006×10⁻⁴ | 0.2366% |

## 5. Conclusion

In this paper, a metrological self-mixing grating interferometer based on the self-traceable chromium atomic deposition grating (SMGI-Cr) is proposed, whose interference phase is traceable to the chromium transition energy level. Based on the SMGI-Cr, one-dimension nanometer displacement along the direction of the grating periodic variation is measured. Compared with a commercial interferometer, the relative displacement is no more than 0.2366%. The SMGI-Cr have the great prospect in the on-site precision measurement without calibration.

## 6. Funding

This work was sponsored by Special Development Funds for Major Projects of Shanghai Zhangjiang National Independent Innovation Demonstration Zone (ZJ2021-ZD-008), National Natural Science Foundation of China (Grant No. 62075165), Shanghai Municipal Science and Technology Major Project (2021SHZDZX0100), Fundamental Research Funds for the Central Universities, and Openning Fund of Shanghai Key Laboratory of Online Detection and Control Technology (ZX2020101).

**Disclosures.** The authors declare no conflicts of interest